\documentstyle[multicol,prl,aps,psfig]{revtex}

\begin{document}

\draft
\title{Quantum inelastic conductance through molecular wires}
\author{H.~Ness and A.J.~Fisher}
\address{Department of Physics and Astronomy, University College London,\\
Gower Street, London WC1E 6BT, U.K.}
\maketitle

\begin{abstract}
We calculate non-perturbatively the inelastic effects on the conductance 
through a conjugated molecular wire-metal heterojunction, 
including realistic electron-phonon coupling.  
We show that at sub-band-gap energies the current is dominated by 
quantum coherent transport of virtual polarons through the molecule.  
In this regime, the tunneling current is strongly increased relative 
to the case of elastic scattering.
It is essential to describe the full quantum coherence of the polaron
formation and transport in order to obtain correct physics.  
Our results are generally applicable to one-dimensional atomic or
molecular wires.
\end{abstract}


\begin{multicols}2

Truly one-dimensional conducting structures, where the confinement
lengths for electrons in two directions are of the order of atomic
diameters, are currently the subject of much experimental work.
Examples include fullerene nanotubes \cite{nanotubes}, organic molecules
bonded to metallic electrodes \cite{reed97}, and dangling-bond (DB) wires
created by scanning tunneling microscope lithography of the H-saturated 
Si(001) surface \cite{DB,DBpeierls}. 
The conductance properties of such structures are of great importance
when one considers their potential as atomic or molecular scale 
electronic devices.
There have been many recent theoretical investigations of electron 
transport in these systems: for example, Joachim {\it et al.} studied 
Xe atom `wires' \cite{pizza97} and through 
organic molecules \cite{CJ_molecwire} using the `elastic scattering 
quantum chemistry' approach; 
idealised one-dimensional atomic wires sandwiched between jellium 
electrodes have been studied using density functional theory
\cite{lang_atwire} and a recursion transfer matrix method
\cite{tsukada98},
while Datta and collaborators studied organic molecules on metal 
electrodes \cite{dattaetal}.  

All the above theoretical work was done within the approximation of
elastic transport, where the electron-phonon ({\it e-ph}) coupling 
plays no role.
But there are good reasons to doubt the validity of this
approximation in one-dimensional atomic or molecular scale systems.
In small systems, the coupling between electron and other
excitations is enhanced. Furthermore, a
one-dimensional metal is generally unstable towards a Peierls
distortion \cite{peierls91}.
Once such a distortion has occurred and
produced a band gap, charges added into the system tend to 
self-localise and cause distortions of the system which
lower the band gap. Such polaronic phenomena have been
studied in conducting polymers for decades \cite{SSHrev};
electron transport usually proceeds via
tunneling into polaron states arising from lattice 
fluctuations \cite{extra_ref}.

Motivated by this physics and the possibility of measuring 
the transport through one-dimensional atomic and molecular 
wires, we report in this letter the first calculations of the 
inelastic electronic transport through an atomic-scale wire including
realistic {\it e-ph} coupling.  
In our calculations, the quantum coherence of the states is fully
retained, i.e., no adiabatic separation between the electronic and phonon
degrees of freedom is made.
We have chosen to study a conjugated molecular chain as an example 
since the {\it e-ph} coupling in such materials is relatively well
understood. 
However the same physical principles (and therefore the qualitative 
features of our results) are widely applicable to all one-dimensional 
atomic and molecular wires.

For energies within the gap, the coherent transport is dominated 
by tunneling. 
Here we study the equivalent of the polaronic phenomena in
conducting polymers for virtual (i.e. tunneling) electron injection.
This has previously been included in very few transport calculations:
the elastic conductivity was found for a conjugated molecule 
containing static solitons \cite{olson98} and the classical response 
of a conjugated 
oligomer to an injected electron wavepacket has been 
considered \cite{Yu99}.
However, the quantum coherence between the electron and 
the lattice was not retained. 
Such coherence is kept in our approach since it is essential to treat 
the tunneling of objects such as polarons or solitons.

Our approach for studying the coupled {\it e-ph} system is
inspired by the Su-Schrieffer-Heeger (SSH) Hamiltonian \cite{SSHrev}, 
in which a tight-binding treatment for the $\pi$-electrons is combined 
with a classical ball-and-spring model for a one-dimensional atomic 
${\rm(CH)}_x$ chain.
In our model, the phonons are treated on a quantum level, the 
{\it e-ph} coupling is linear, and the electronic part is 
expressed in terms of the one-electron eigenstates, labelled by $n$.  
The Hamiltonian is
\begin{equation}
H=\sum_n\epsilon_n c^\dag_n c_n
+\sum_q\omega_q a^\dag_q a_q
+\sum_{q,n,m}\gamma_{qnm}(a^\dag_q+a_q)c^\dag_n c_m,
\label{Hamilt}
\end{equation}
where $c^\dag_n$ ($c_n$) creates (annihilates) an electron in the
$n$-th electronic state with energy $\epsilon_n$ and $a^\dag_q$ ($a_q$) 
creates (annihilates) a phonon in mode $q$.
This is a general Hamiltonian, describing coupled {\it e-ph}
systems going beyond the Holstein model.
 
In this work, the electronic energies $\epsilon_n$, the phonon 
frequencies $\omega_q$ and the coupling matrix elements $\gamma_{qnm}$ 
are obtained from the SSH Hamiltonian for an isolated molecular 
chain \cite{note_paramSSH}.
The one-electron eigenstates and energies $\epsilon_n$ are calculated 
self-consistently with the atomic distortions for a given number of 
atoms $N_{\rm a}$ and $\pi$-electrons $N_{\rm e}$. 
The ground state of the neutral ($N_{\rm e}=N_{\rm a}$) dimerised chain 
is our reference system, from which the phonon
modes and frequencies are obtained by perturbation theory \cite{chao85}. 
The energy change, taken to second order in the atomic 
displacements $x_i$, gives the dynamical matrix from which phonon 
eigenmodes $V_q(i)$ and frequencies are obtained.  
We have checked that this second-order expansion of the elastic energy 
is sufficient to describe the formation of a static polaron in the 
chain.
The {\it e-ph} coupling matrix elements are calculated from 
the variation of the electronic Hamiltonian due to 
phonon displacements.  
They are transformed from a real-space representation into the electron 
eigenstate representation to get the $\gamma_{qnm}$ elements in 
Eq.(\ref{Hamilt}).

\begin{figure}[p]
\psfig{figure=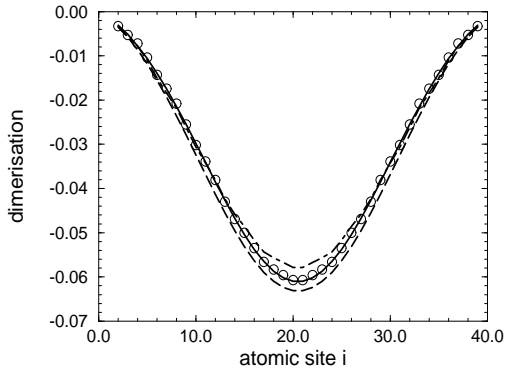,width=8cm,height=5cm}
\narrowtext
\caption{
\label{fig1}
Dimerisation pattern $d_i=(-1)^i(x_{i+1}-2x_i+x_{i-1})$ (in \AA) 
induced by an additional electron in the isolated wire 
($N_{\rm a}=40$, $N_{\rm e}=41$): classical SSH model 
({\it dashed line}); half electronic spectrum and all classical 
phonons ({\it solid line}); $N_{\rm ph}=6$ (lowest optical modes) 
classical phonons ({\it circles}), $N_{\rm ph}=6$ (same modes) 
quantum phonons with $n_{\rm occ}^{\rm max}=6$ ({\it dot-dashed line}).}
\end{figure}

We choose the eigenstate representation because we wish to consider
the action of $H$ on the $(N_{\rm a}+1)$-electron (or $(N_{\rm
a}-1)$-electron) Hilbert space obtained by adding (or removing) one
electron to (or from) the ground state of the neutral chain. 
It is most straightforward to project out the addition of an electron 
(or hole) to the already occupied (or empty) states
if we work entirely with the eigenstates. 
Then the $n,m$ sums in Eq.(\ref{Hamilt}) run over the occupied 
valence band states (for hole transport) or over the unoccupied 
conduction band states (for electron transport).
We note that it is essential to keep the $N_{\rm a}$ electrons of the 
reference system implicitly in the calculation since they drive the 
Peierls distortion in the system. 
We justify the inclusion of only $(N_{\rm a}\pm1)$-electron states in 
the transport by noting that the mean time between electron (or hole) 
passages ($\approx 10^{-7}\,\rm s$ for a current of 1\,pA) is much 
longer than a typical residence time in the molecule 
($\approx 10^{-15}\,\rm s$ for a molecule strongly coupled to the
electrodes \cite{notetime}).
Furthermore, we have verified that the static lattice distortions due 
to charging are well reproduced by considering only one half 
of the electronic spectrum (Fig. \ref{fig1}).

The degree of dimerisation is conveniently represented by the 
staggered difference $d_i$ between adjacent bond lengths.
Figure \ref{fig1} shows the $d_i$ pattern induced when one 
additional electron is injected into the neutral isolated chain. 
As expected, we find a reduction of dimerisation, corresponding to a 
polaron-like defect in the middle of the chain. 
For short chains ($N_{\rm a}$=40 for example), the difference of
the dimerisation calculated with (a) the full electronic spectrum and
an exact classical treatment of the lattice (SSH model), and (b) half 
of the electronic spectrum and a harmonic treatment of the lattice, 
does not exceed 7\% \cite{note_atdisto}. 
We also checked that only the longest wavelength optical phonon modes 
\cite{notephE} contribute to the polaron distortion.

The basis states for the Hamiltonian Eq.(\ref{Hamilt}) are expressed
as $\vert n,\{n_q\}\rangle$ where $n$ indexes the one-electron
eigenstates and $\{n_q\}$ represents the set of occupation numbers for
the different phonon modes $q$.
The elastic states $\vert
n,\{0\}\rangle$ have no excited phonon modes and the inelastic
states $\vert n,\{n_q\}\rangle$ have phonon modes $q$ excited with
$n_q=1,2,3,..,n_{\rm occ}^{\rm max}$.
In all the calculations, the dynamical (i.e. energy dependent) 
correlation between the electron and the phonons is fully kept.
Even for a relatively short wire the size of the complete basis set quickly 
becomes intractable for reasonable numerical calculations.  
However using a limited but sufficient number of phonon modes to describe 
the polaron formation permits us to treat a range of chain lengths with 
reasonable computing times. 
Furthermore, our results show that only a few excitations in each phonon 
mode are needed to give converged results.

Knowing $\epsilon_n$, $\omega_q$ and $\gamma_{qnm}$ for a realistic 
description of the wire, we then calculate the transport properties through 
the electrode/wire/electrode junction. 
The conjugated chain is connected to two semi-infinite 1D metallic 
electrodes of band width 6 eV via the use of embedding potentials 
$\Sigma_L$ and $\Sigma_R$ \cite{note_tLR} .
The propagation of an electron or hole through the junction is treated 
as a scattering problem with many channels \cite{bonca95,bonca97,ness98}. 
Full scattering boundary conditions are applied for an incident electron 
or hole plane wave with the wire in a given vibrational state. 
For the results given in this paper, the chain configuration before 
scattering is the ground state (no phonons present).  
The solution of the scattering problem can be obtained by solving the 
following linear system
\begin{equation}
\left[
E-H-\Sigma_L(E)-\Sigma_R(E)
\right]
\vert \alpha(E)\rangle
=
\vert s(E)\rangle\ ,
\label{linsys}
\end{equation}
where $\vert s(E)\rangle$ is the source term, representing the injected
carrier at energy $E$.  The ket $\vert \alpha(E)\rangle$ is
the wave function of the scattering state; it is represented as an
expansion over the basis set $\vert n,\{n_q\}\rangle$.  
We can derive the transmission and reflection coefficients for the 
different channels, the currents and the average atomic displacements
from the solution \cite{bonca97,ness98} of Eq.(\ref{linsys}) obtained
by a conjugate gradient technique.  

To check the validity of our quantum approach, 
we first calculate the mean atomic displacements $\langle x_i\rangle$ 
induced by adding an electron to an isolated chain in the ground 
state \cite{note_defxi}. 
The corresponding dimerisation pattern, also shown in Fig.\ref{fig1},
reproduces the results obtained with classical phonons.
The dip in the dimerisation is slightly shallower in the fully 
quantum case; this may result from quantum delocalisation 
of the polaron state.

\begin{figure}[p]
\psfig{figure=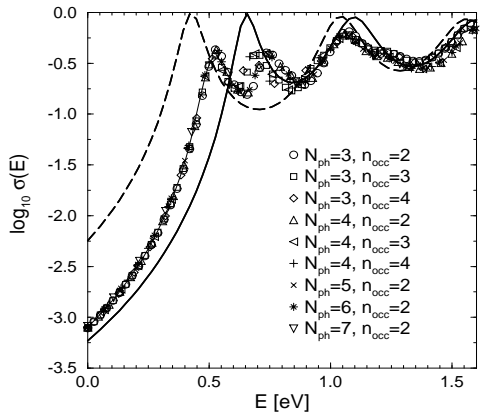,width=8cm,height=5.5cm}
\narrowtext
\caption{
\label{fig2}
Differential conductance $\sigma(E)$ (in units of $2e^2/h$)
through the wire ($N_{\rm a}=40$). 
$\sigma(E)$ with inelastic scattering is 
represented by the symbols corresponding to different $N_{\rm ph}$ 
optical modes and $n_{\rm occ}^{\rm max}$.
The thin solid line is a guide for the eye.
The elastic conductance through the static neutral chain 
({\it solid line}) and charged chain containing a static polaron 
({\it dashed line}) is also shown
($E=0$ is at mid-band-gap).}
\end{figure}

\begin{figure}[p]
\psfig{figure=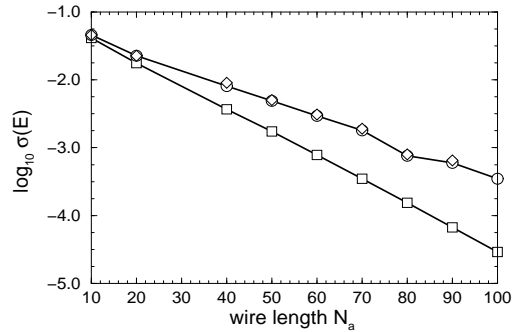,width=7.5cm,height=4.5cm}
\narrowtext
\caption{
\label{fig3}
Wire length dependence of $\sigma(E)$ (in units of $2e^2/h$)
for an injection energy $E=0.3$ eV inside the tunneling gap.
The inelastic calculations are obtained for $N_{\rm ph}=4$ lowest 
optical modes, $n_{\rm occ}^{\rm max}=2$ ({\it circles}) and 
$n_{\rm occ}^{\rm max}=3$ ({\it diamonds}).
The elastic conductance through the static neutral chain is also
shown ({\it squares}).} 
\end{figure}

We now turn to calculations on a current-carrying wire between two
electrodes.  Figure \ref{fig2} shows the differential conductance
$\sigma(E)$ of the junction.  
Calculations are performed with different numbers of phonon modes and 
with different maximum occupation numbers $n_{\rm occ}^{\rm max}$.  
We also calculate the elastic conductance through a static neutral 
chain  and a charged chain containing a static polaron in the centre.  
One important result of our approach is that the inclusion of inelastic 
scattering strongly enhances the current in the tunneling regime compared 
to the case of elastic scattering.  
The transport becomes increasingly dominated by resonant tunneling through 
virtual polaron states as the injection energy approaches the top of the 
band gap.  The first resonance in $\sigma(E)$
occurs at an injection energy very close to the charging energy
$E_{\rm ch}$ of the isolated molecular chain (for $N_{\rm a}=40$, we
calculate $E_{\rm ch}=0.53$ eV).  
As expected, for any injection energy within the tunneling gap, the 
current decreases exponentially as the length of the wire 
increases (Fig. \ref{fig3}) for both elastic and inelastic calculations.  
The absolute value of $\sigma(E)$ is much larger when the 
{\it e-ph} coupling is included, and the apparent band gap
(related to the slope of the curve) is smaller. Such effects become 
more important for longer molecular wires ($N_{\rm a}\geq 100$).
For shorter wires ($N_{\rm a}\leq 10$), both calculations seem to
converge; the tunneling process is too fast for the lattice
to respond significantly to the presence of the electron,
so the effects of {\it e-ph} coupling are less 
pronounced \cite{note_nashort}.

\begin{figure}[p]
\psfig{figure=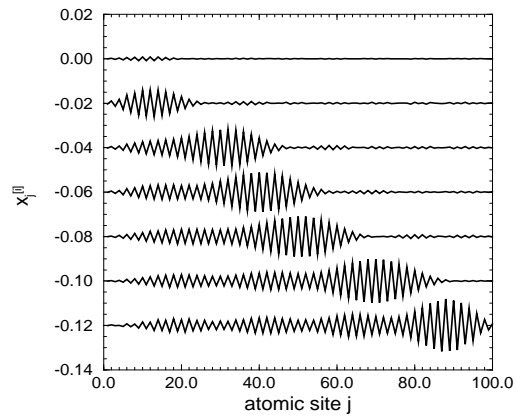,width=8cm,height=5.5cm}
\narrowtext\caption{\label{fig4}
Induced atomic displacements $x_j^{[i]}$ (in \AA) due to a 
tunneling electron (injection energy at mid-gap) 
for a $N_{\rm a}=100$ chain.
The curves are shifted vertically (by 0.02) for clarity.
The electron-projection $\hat{P}_i$ sites are $i$=1,11,31,41,51,71,91
for the top to the bottom curve respectively. 
Calculations are performed for $N_{\rm ph}=8$ optical modes 
and $n_{\rm max}^{\rm occ}=2$.}
\end{figure}

Finally, we present in Figure \ref{fig4} the quantum averaged
induced atomic displacement $x_j^{[i]}$ of an atomic site $j$ when the
tunneling electron is found on site $i$ \cite{note_defxij}.  
These results are characteristic of phonon fluctuation-assisted 
tunneling via the dynamical formation of a virtual polaron inside 
the chain.

The bond alternation, and hence the band gap, are locally suppressed 
around the tunneling electron. 
This distortion accompanies the electron through the chain,
leaving a `wake' of distortion behind it.  Such a persistent response
of the lattice is not surprising in the absence of damping \cite{bonca97}.  
The displacements are slightly less than those for an isolated chain
(not shown) because the lattice cannot respond fully to the tunneling 
electron.
This effect manifests itself in the calculated conductance
(Fig. \ref{fig2}); when tunneling, the current in the full quantum
coherent calculations is less than the current that flows through a 
static chain where the polaron is `frozen in'.  
This is because (a) the effective distortion in the quantum calculations 
is slightly smaller, and (b) the frozen-in polaron already exists so
no elastic energy is needed to create it.
Only fully quantum coherent calculations give qualitatively correct 
physics; it cannot be obtained from elastic scattering calculations
through static chains (with or without disorder) or using a classical 
lattice.

To conclude, we have presented the first calculation of inelastic
effects on the conduction through molecular wires by explicitly 
considering realistic electron-phonon coupling.  
We have shown that the quantum coherent dynamical formation and transport
of virtual polarons inside the molecular wire is crucial to describe the 
electronic transmission properties of such a system, and leads to a 
strongly increased current in the tunneling regime.  
This is the small-bias regime in which molecular wires will normally 
operate.
Therefore, whenever tunneling conduction occurs in these or other
one-dimensional atomic-scale wires \cite{note_1datw} and most
probably in conducting polymer wires \cite{extra_ref},
it will be by the mechanism identified here. 

We thank A.M. Stoneham, L. Kantorovich, J. Gavartin and A. Shluger
for useful comments,
S. Shevlin for contributions to the present study.
This work was supported by the U.K.\ EPSRC.

\narrowtext

\end{multicols}

\begin{references}
\bibitem{notetime}
The residence time can be estimated from the resonance widths in
the transmission. 
We also calculated the B\"uttiker-Landauer tunneling time 
$\tau^{\rm BL}$ as in Ref. \cite{ness98}. 
We find $\tau^{\rm BL}\approx$ 2 to 4 fs for $N_{\rm a}$=40 
and $\tau^{\rm BL}\approx$ 8 to 20 fs for $N_{\rm a}$=100.
\bibitem{note_atdisto} 
For longer chains ($N_{\rm a}\approx 100$ for example), the
difference in the dimerisation increases but is still in the
range 15-17\%.  This may be because the band gap decreases
for longer chains.  However, the atomic distortions due to charge
injection are still well reproduced.
\bibitem{note_tLR}
The leads are 1D atomic chains with hopping 
integrals $\beta_{L,R}$=3 eV.
The matrix coupling elements between the end atoms of the chain and
the first atoms of the left and right leads are
$t_{L,R}$=2.2 eV.
The main results shown are not strongly dependent on 
those parameters.
\bibitem{note_defxi}
The mean atomic displacement $\langle x_i\rangle$ is obtained from
$\langle x_i\rangle=\sum_q V_q(i)\langle X_q\rangle$ where 
$\langle X_q\rangle$ is the mean displacement of the phonon mode $q$: 
$X_q=\sqrt{{\hbar}/{2m\omega_q}}(a^\dag_q+a_q)$.
\bibitem{note_nashort}
Also, for short wires, the polaron displacement cannot be accommodated.
\bibitem{note_defxij} 
The averaged atomic displacement $x_j^{[i]}$ of atom $j$ when the 
electron is found on site $i$ is obtained from 
$x_j^{[i]}=\sum_q V_q(j) X_q(i)$ where $X_q(i)$ is the averaged 
displacement of mode $q$ when the electron is at site $i$, defined 
by $X_q(i)=\langle\hat{P}_i\hat{X}_q\hat{P}_i\rangle/
\langle\hat{P}_i\rangle$. 
$\hat{P}_i$ is the electron wave function projector onto the atomic 
site $i$.
\bibitem{note_1datw}
Another system to which this mechanism will be relevant is
the DB wire on the H-passivated Si(001) surface 
for which Peierls distortion occurs \cite{DBpeierls,DBpizza99} and 
where a polaron-like defect is formed due to carrier 
injection \cite{future}.
\end{references}
\end{document}